\documentclass{article}

\usepackage{arxiv}

\usepackage[utf8]{inputenc} 
\usepackage[T1]{fontenc}    
\usepackage{hyperref}       
\usepackage{url}            
\usepackage{booktabs}       
\usepackage{amsfonts}       
\usepackage{nicefrac}       
\usepackage{microtype}      
\usepackage{lipsum}
\usepackage{graphicx}
\title{Mitigation of carrier induced optical bistability in silicon ring resonators}

\author{
  Vadivukkarasi Jeyaselvan \\
  Centre for Nano Science and Engineering\\
  Indian Institute of Science\\
  Bangalore\\
  \texttt{shankarks@iisc.ac.in} \\
   \And
  Shankar~Kumar~Selvaraja \\
  Centre for Nano Science and Engineering\\
  Indian Institute of Science\\
  Bangalore\\
  \texttt{vadivu@iisc.ac.in} \\
}

\begin{document}
\maketitle
\begin{abstract}We present a detailed study of electrical and optical generated free carrier on the spectral characteristics of a silicon microring modulator. The spectral distortion generated due to thermal and free carriers is presented, and the mechanism for mitigation is also presented. We

 observed that two-photon induced nonlinearity could be addressed by operating the modulator at suitable bias points. Furthermore, by applying small-signal drive the spectral distortion can be restored. We also present the effect of optical power and drive signal limit on the spectral characteristics. The study allows one to identify suitable device performance and operating conditions to utilize silicon ring modulator for optical signal processing.

\end{abstract}

\section{Introduction}

High-Index contrast silicon-on-Insulator (SOI) allows tight confinement of light in a sub-micron cross-section that supports compact light guiding and manipulation. Besides, SOI provides a promising integration platform for CMOS electronics and photonics~\cite{soref1986all,lipson2005guiding}. The optical confinement in the waveguides has been exploited for various applications~\cite{horvath2003optical, cheng2018recent,perez2018field}. A further enhancement in power-density is achieved by using resonance structures such as ring resonators, Bragg cavity and photonic crystal nanocavities~\cite{zhang2014high}. There have been recent advances in obtaining high cavity quality factors (Q-factor) as high as $10^{6}$~\cite{bogaerts2012silicon} in a smaller footprint. Though higher optical confinement in Si results in compact device geometries, it leads to nonlinear effects such as two-photon absorption (TPA)~\cite{priem2005optical}, carrier induced optical bi-stability~\cite{xu2006carrier}, thermally induced nonlinearities~\cite{ilchenko1992thermal}, free carrier absorption (FCA)~\cite{claps2004influence}, and free-carrier dispersion (FCD). These optical nonlinearities are reported to occur at optical power as low as 0.277~mW~\cite{priem2005optical}.  The carriers created by FCA and TPA leads to FCD resulting in a blue-shift in the spectral response of the devices. Some of the generated carriers undergo interband and intraband relaxation that leads to thermal-phonon creation. Since Si has positive thermo-optic coefficient, heating cause red-shift in the spectral response of the devices. The characteristic time constant of the thermal effect is relatively slow ($\mu$s) while the FCD process is much quicker ($ns$). The interplay between these effects has lead to the realization of all-optical modulators ~\cite{pandey2018all,xu2006carrier,almeida2004optical}. Though the nonlinearity is exploited for a few applications, it causes undesirable performance degradation in many carrier dependent or high-power Si devices.  

Ring resonators find application in modulation and dense wavelength filters~\cite{chen2015comb}. Due to high-power density in the ring cavity, the cavity quickly moves into nonlinear behaviour even at lower input power. As mentioned earlier, interplay between the carrier and thermal dispersion results in drift and degradation in device performance. Recently, there has been a demonstration of frequency comb generation using Si ring modulator~\cite{demirtzioglou2018frequency} where the nonlinear effects in the ring response distort the symmetricity in comb lines. Besides, these applications also demand a stable frequency operation. A wide range of thermal-induced phenomena such as hysteretic wavelength response~\cite{xu2006carrier} and oscillatory instability~\cite{ilchenko1992thermal} were also experimentally reported. Exploitation of the linear and nonlinear phenomena request an understanding of optical and electrical dependences on the process. Furthermore, a mechanism to mitigate the undesirable nonlinear process is required.
 
In this article, we report a comprehensive experimental study and analysis of optical power and electrical signal dependent nonlinear effects in Si ring cavity. We also demonstrate mitigation of the undesirable spectral distortion of the ring modulator by using an external electric field across the waveguide. Using bias tunable AC signals, we study the effect of thermal and free-carrier impact on the transmission spectrum at various drive frequencies.

\section{Theory} \label{sec:theory}

\begin{figure}[tp]
\centering
\includegraphics[width=\linewidth]{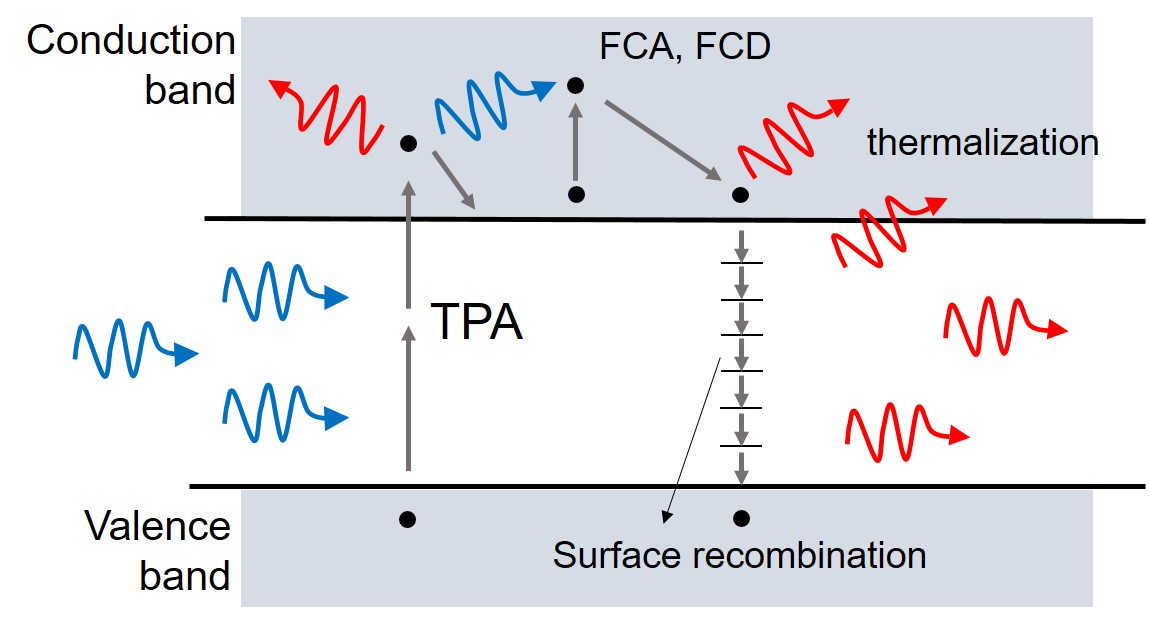}
\caption{Schematic of nonlinear interactions in silicon for wavelengths around half the band gap.}
\label{fig:Theory}
\end{figure}

The bandgap of silicon is 1.12~eV, which makes it transparent for wavelengths beyond 1100 nm. However, when illuminated with photon energy below the bandgap energy ($>1100~nm$) and sufficient optical power could result in multi-photon absorption and associated nonlinear process as illustrated in Fig.~\ref{fig:Theory}. The high power density of a propagating mode at 1550 nm wavelength band in a Si waveguide could result in TPA process and subsequent change in refractive index proportional to the light intensity (Kerr effect). Since Kerr coefficient is very small in silicon, the refractive index change due to the Kerr effect is negligible. The free carriers excited through the TPA process contribute to FCA that results in refractive index reduction referred to as FCD~\cite{soref1986all}. The carriers generated primarily relax to ground state either through interband or intraband relaxation creating thermal phonon. The heating due to thermal phonons gives rise to positive change in refractive index resulting in red-shift in the spectral response. The FCD-induced and thermally-induced refractive index change compete with each other, and these effects exhibit different characteristic time constants~\cite{pernice2010time}. The largest of time constants associated with these processes determine the resultant refractive index change~\cite{priem2005optical}.

The undesirable carriers could be swept by applying an electric field across the waveguide~\cite{jalali2006raman}. The field across a waveguide or a junction in an active device could reduce the undesirable spectral shift due to refractive index drift. In this paper, we use a Si PN junction type micro-ring modulator as a test device for the demonstration of carrier mitigation.

\section{Optical bistability in silicon ring modulator}
\label{sec:bistable}
\subsection{Test device design}

\begin{figure}[htb]
\centering
\includegraphics[width=\linewidth]{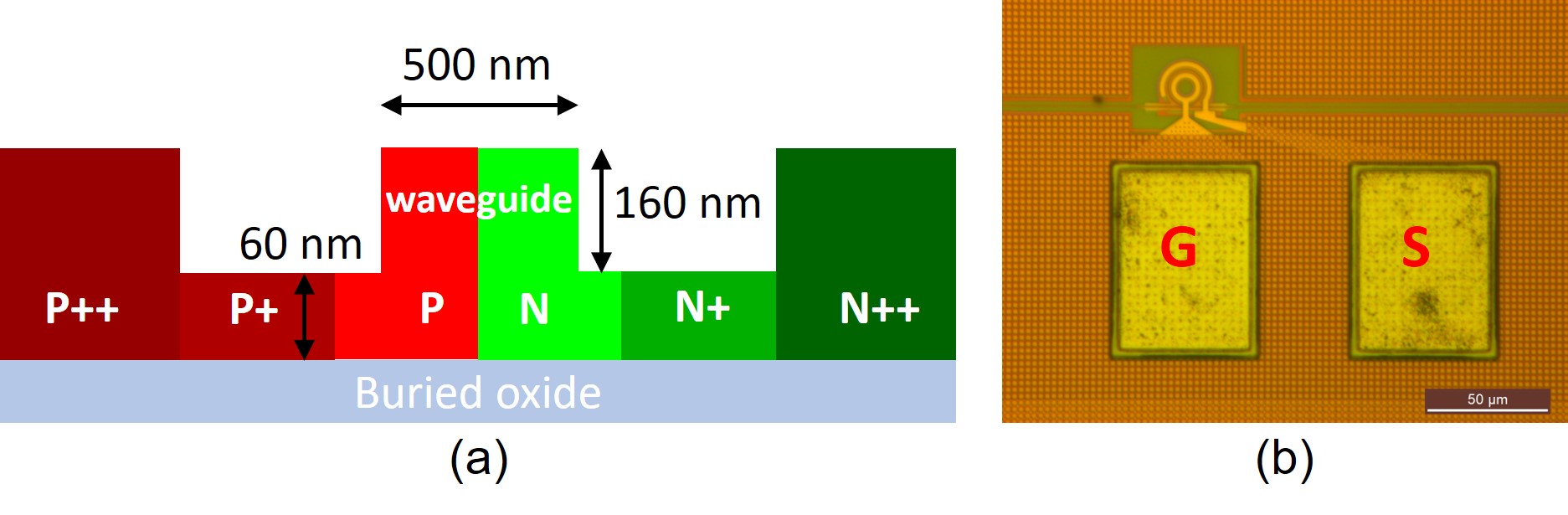}
\caption{(a) Cross-section schematic of a PN junction across the waveguide and (b) microscope image of a fabricated silicon ring modulator.}
\label{fig:CS}
\end{figure}

We use high-speed depletion-mode silicon ring modulator with an electro-optic bandwidth of 25 GHz~\cite{MPW} as a test device. Fig.~\ref{fig:CS}(a-b) shows the schematic cross-section of the PN junction across the waveguide and the microscope image of the ring modulator. The device is fabricated on an SOI substrate with 220~nm thick silicon device layer on 2~$\mu$m buried oxide.  The rib waveguide is designed with a width and height of 500~nm and 160~nm, respectively. Grating fiber-chip couplers were used to couple light with a coupling efficiency of 5 dB/coupler. The gap between the waveguide and the ring is 180~nm. The radius of the ring is chosen to be 7.5~$\mu$m.  

\subsection{Optical bistability in ring resonator}

\begin{figure}[hb]
\centering
\includegraphics[width=0.65\linewidth]{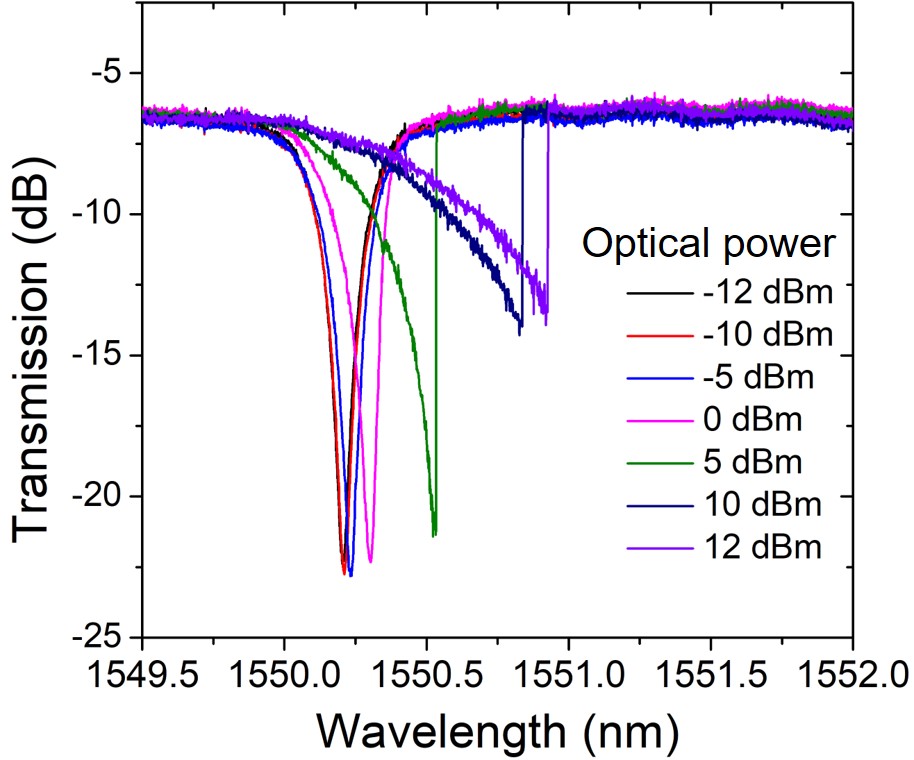}
\caption{Transmission spectra of a ring modulator with various optical power.}
\label{fig:Txn}
\end{figure}

\begin{figure}[htb]
\centering
\includegraphics[width=\linewidth]{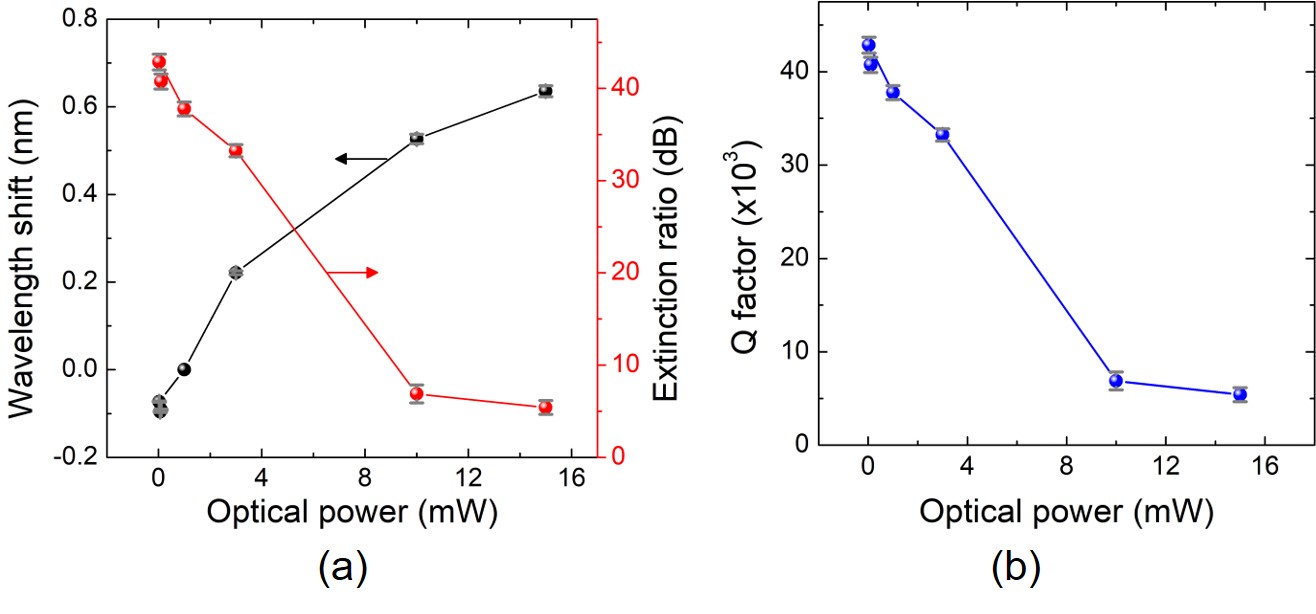}
\caption{Effect of input optical power on (a) Wavelength shift, (a) Extinction ratio and (b) Q-factor of a silicon ring modulator.}
\label{fig:Txn_ext}
\end{figure}

Optical characterization was done using a tunable wavelength laser (1500-1630 nm) and a photodetector. All characterization is done at a stable temperature (30$^{o}$C) using a temperature stabilized measurement stage. The transmission spectrum of the ring resonator is measured as a function of input optical power; 0.063-15.85~mW~(-12 to +12~dBm), and is shown in Fig.~\ref{fig:Txn}. It is clearly observed that the resonance shape, wavelength, and Q-factor shifts with increasing optical power. Asymmetricity in lineshape~\cite{almeida2004optical,carmon2004dynamical} is observed when the optical power increases and it occurs for optical power as low as 0.3~mW~(-5.23~dBm). As discussed in Sec.~\ref{sec:theory}, the observed bistability is due to FCD and recombination associated thermally-induced refractive index change. Since the power density in an optical cavity depends on the Q-factor, it is essential to understand the effect of input power on the carrier dynamics in the ring cavity. The minimum power required to drive the cavity into nonlinear or bi-stability regime depends on the finesse of the cavity; cavity with larger finesse would have lower power threshold for nonlinear behaviour. The undesirable spectral behaviour has direct implication on some of the key applications, such as optical links and optical frequency comb generation~\cite{demirtzioglou2018frequency}. Fig.~\ref{fig:Txn_ext}(a-b) shows the effect of optical power on the cavity resonance. As the injected power increases, we observe a positive- or red-shift in the resonance wavelength. However, at lower power we observe a negative or blue-shift in the resonance wavelength (Fig.~\ref{fig:Txn_ext}(a)). The optically generated carriers in the cavity increase with increasing power. Beyond the threshold power of 1~mW, the thermally induced refractive index change dominates such that we observe a red-shift. The FCA also increases with increasing power that primarily contribute to absorption loss which is evident from the reduction in extinction  (Fig.~\ref{fig:Txn_ext}(a)) and cavity Q-factor (Fig.~\ref{fig:Txn_ext}(b)).

\begin{figure}[htb]
	\centering
	\includegraphics[width=0.65\linewidth]{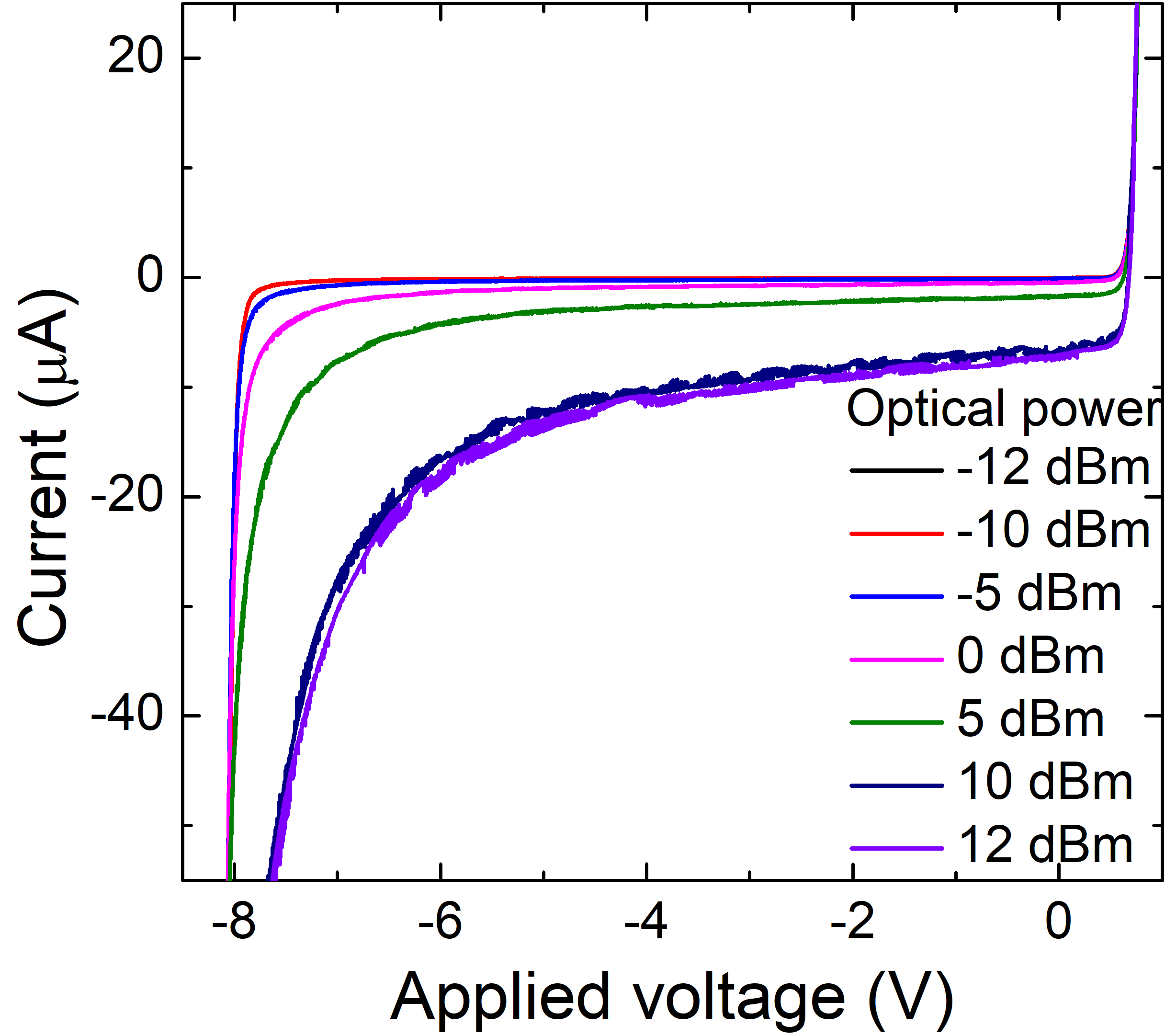}
	\caption{Current-voltage characteristics of the PN junction across the waveguide at various input optical powers.}
	\label{fig:IV_PN}
\end{figure}

The DC electrical characteristics of the ring cavity with PN junction is measured at various input optical powers. Fig.~\ref{fig:IV_PN} shows the current-voltage (I-V) characteristics of the PN junction diode present across the waveguide. The I-V characteristics in the absence of an optical field is similar to simple PN diode. However, with increasing optical power the characteristics are similar to a solar cell~\cite{lindholm1979application}.  Unlike an ideal diode, a reverse breakdown ($V_{BR}$) is observed at ~-8 V. With increasing optical power,  $V_{BR}$ decreases which indicate an increase in carrier concentration in the PN junction. Beyond 10 dBm, the carrier concentration saturates in the cavity. The observed saturation in $V_{BR}$ can correspond to the reason for saturation in the extinction and Q-factor of the ring with increasing optical power (Fig.~\ref{fig:Txn_ext}). The optically generated carriers induce spectral distortion and bistability in the ring response. These generated carriers can either be drifted across the PN junction using an electric field in reverse-bias mode~\cite{dimitropoulos2005lifetime} or can be allowed to recombine with the carriers in the waveguide core in the forward bias regime.

\section{Results and Discussion}

\begin{figure}[htb]
	\centering
	\includegraphics[width=\linewidth]{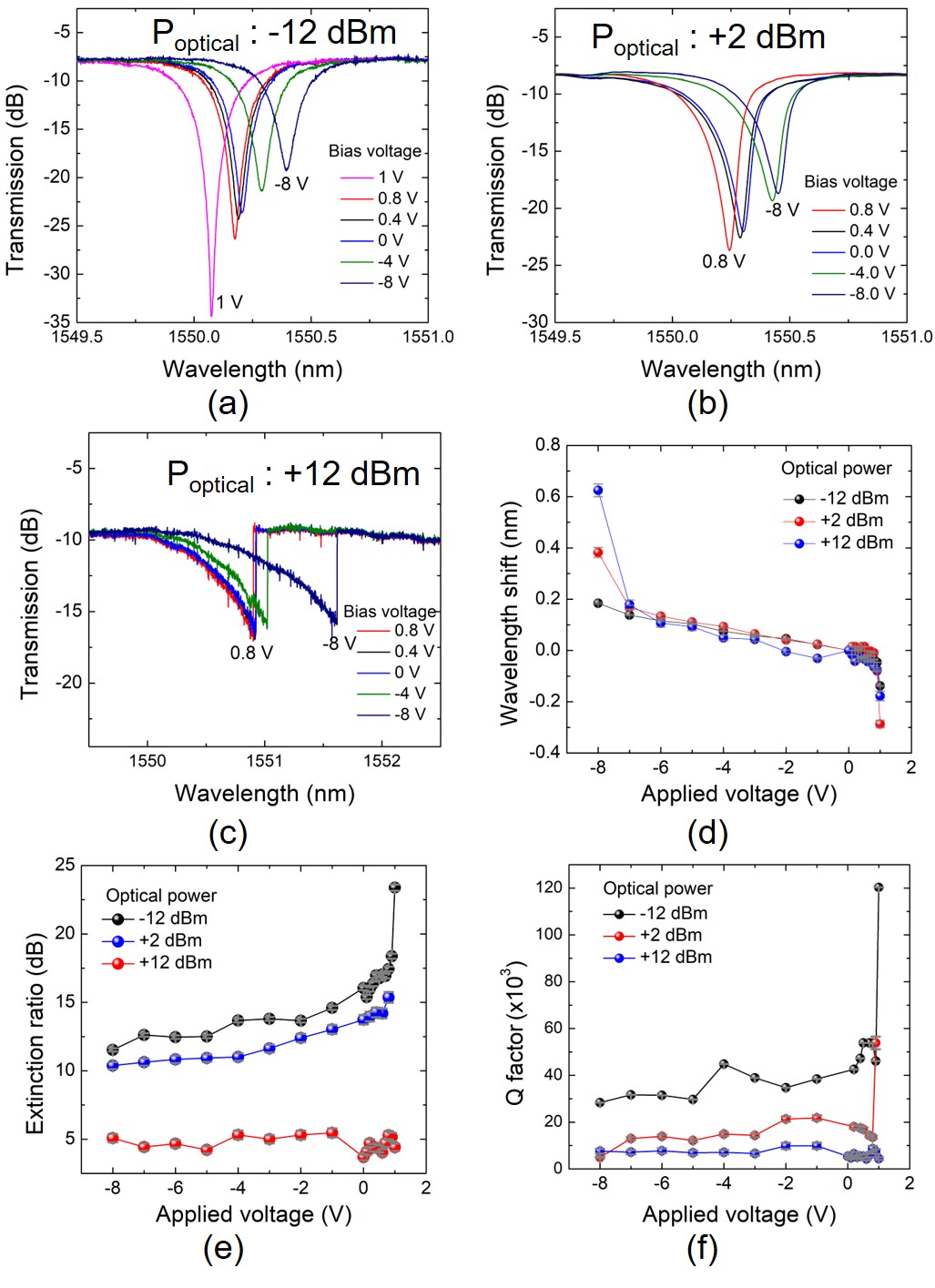}
	\caption{ Effect of input optical power and DC bias on the transmission spectra of the ring resonator, (a)~Optical power of -12~dBm, (b)~Optical power of +2~dBm, (c)~Optical power of +12~dBm. Spectral characteristics of the ring at various optical power and bias (d)~Wavelength shift, (e)~Extinction ratio and (f)~Q-factor.}
	\label{fig:Txn_voltage}
\end{figure}

We analyze the spectral characteristics of the ring with various DC-bias voltages and optical powers. Fig.~\ref{fig:Txn_voltage}(a-c) shows the effect of applied DC-bias on the transmission spectral characteristics of the device at various input optical power of~-12, +2, and +12~dBm. At reverse-bias voltages, the carriers are depleted across the waveguide resulting in a red-shift in the spectrum while forward-bias causes blue-shift in the spectrum~\cite{soref1987electrooptical}. It can be observed from Fig.~\ref{fig:Txn_voltage}(a) that at a lower input power of -12~dBm, irrespective of the applied bias, the shape of the resonance; Lorentzian profile, is preserved. However, at higher optical power of +12~dBm (Fig.~\ref{fig:Txn_voltage}(c)), the spectral characteristic is highly asymmetric exhibiting bistable behaviour. At ~+2~dBm of input power, asymmetry starts to evolve at higher reverse-bias voltages (Fig.~\ref{fig:Txn_voltage}(b)). As mentioned earlier, the power threshold for nonlinear response depends on the nature of the cavity; cavity-loss and Q-factor, thus fabrication technology and device design also play an important role.

It can be observed from Fig.~\ref{fig:Txn_voltage}(a-b) that the spectral asymmetry is higher at reverse bias compared to forward bias. This is because at higher optical generation rate, in the reverse-biased regime, the electric field built due to drift electron and holes oppose the applied electric field. At +12~dBm irrespective of bias (Fig.~\ref{fig:Txn_voltage}(c)) we observe spectral asymmetry due to the high concentration of optically generated carriers beyond compensation, which is confirmed by stable extinction across the bias voltages. Hence at higher optical powers, reverse-bias PN junction becomes ineffective in carrier compensation \cite{dimitropoulos2005lifetime}. However, beyond the compensation limit, irrespective of the bias polarity the spectral characteristic is unaffected as observed from Fig.~\ref{fig:Txn_voltage}(c).

In addition to the spectral profile change, the resonance wavelength also shifts. Fig.~\ref{fig:Txn_voltage}(d) shows the wavelength shift as a function of applied bias voltage at various input optical powers. Irrespective of the optical power and reverse-bias voltage we observe similar wavelength shifts for voltages below $V_{BR}$. However, at the onset of reverse breakdown, the wavelength shift increases with increasing optical power due to the avalanche proportional to the optically generated carriers. The wavelength shift increases exponentially in the forward bias regime with applied bias resembling the current-voltage characteristics of a PN junction (Fig.~\ref{fig:IV_PN}). The red ($+ve$) and blue ($-ve$) in the wavelength shift shows the carrier concentration change in the junction due to reverse- and forward-bias, respectively.

 Fig.~\ref{fig:Txn_voltage}(e-f) shows the extinction ratio and Q-factor extracted from the transmission spectrum of the ring modulator at different optical power and bias voltages. We observe the effect of an electric and optical generated carrier on the extinction and Q-factor. At higher optical injection (+12 dBm), due to carrier saturation both extinction and Q-factor is unaffected by applied bias voltages within forward and reverse breakdown limit. However, for lower injection, we observe increase in both extinction and Q-factor due to applied bias. With increasing reverse-bias voltage, we observe a reduction in extinction with marginal reduction in Q-factor. However, in the forward regime, we observe increase in extinction and Q-factor. The observed trend in the forward bias is due to increase in carrier concentration resulting in operation regime change from over-coupling to critical coupling. Hence, in this work, PN junction is operated at forward bias mode where the depletion region is small and more carriers are diffused into the junction to allow recombination with the generated optical carriers.

\begin{figure}[tb]
\centering
\includegraphics[width=\linewidth]{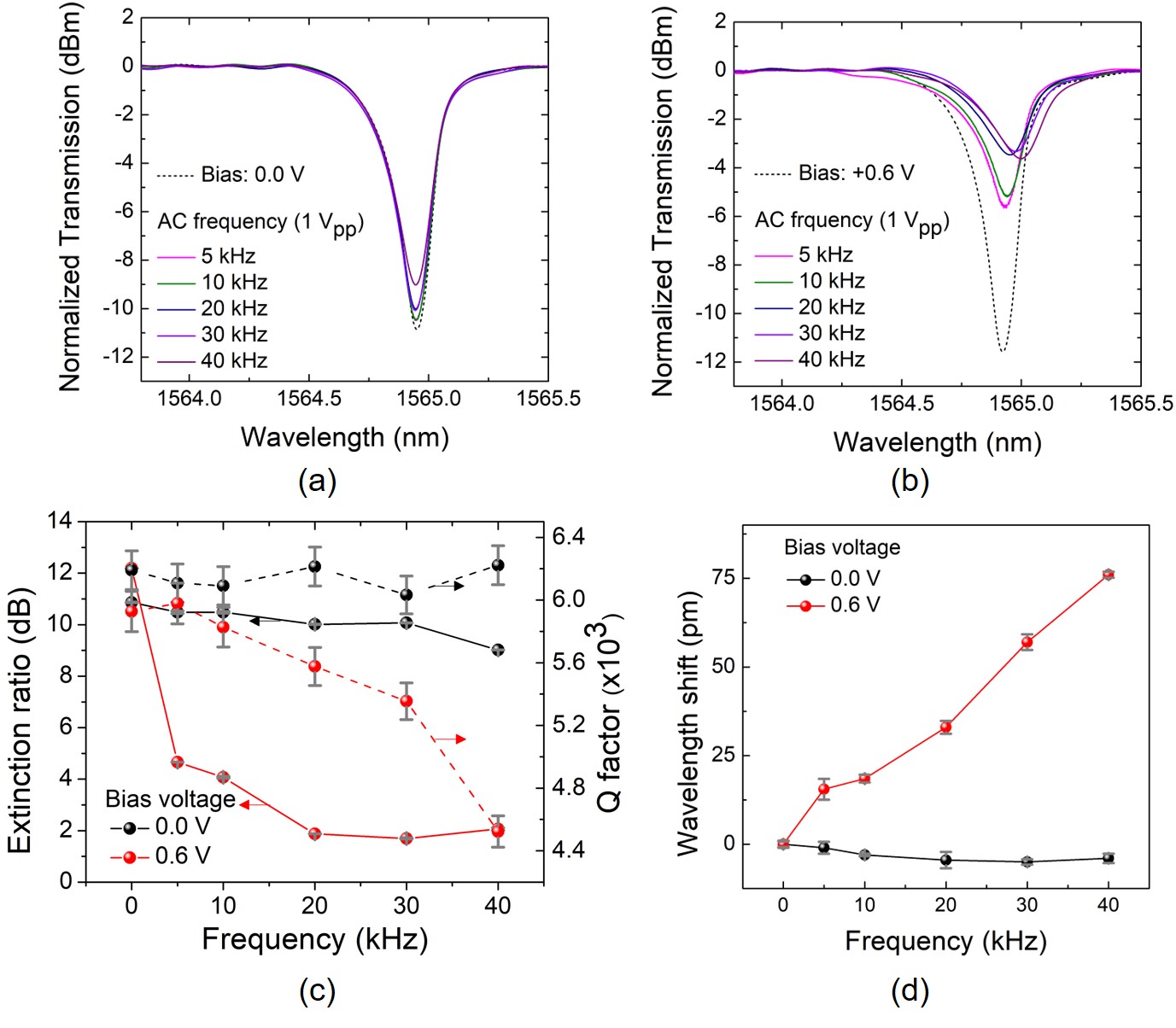}
\caption{Transmission spectra of the ring resonator at various small-signal AC feed (5-40 KHz) at (a) Bias voltage of 0~V and (b) Bias voltage of +0.6~V. Effect of AC feed at 0~V and +0.6 V bias on the (c)~Extinction ratio and Q-factor, and (d) Wavelength shift at an input optical power of +2~dBm.}
\label{fig:DiffBiasVoltage_RF1Vpp}
\end{figure}

The dynamic characteristics of the ring modulator are studied by applying a small-signal AC at various frequencies ($<40 kHz$). The small-signal assists in sweeping the generated optical carriers from the waveguide. A positive-bias voltage is chosen as it reduces the asymmetricity of the resonance. Fig.~\ref{fig:DiffBiasVoltage_RF1Vpp}(a-b) shows the transmission spectrum of the ring modulator for different frequencies at a bias voltage of 0~V and +0.6~V for a fixed peak-peak voltage ($V_{pp}$) of 1~V at an input optical power of +2~dBm. The spectral characteristics show negligible change with frequency at 0~V, however, at +0.6~V we observe resonance shift as well as reduction in the extinction. Fig.~\ref{fig:DiffBiasVoltage_RF1Vpp}(c-d) depicts the extracted extinction, Q-factor, and wavelength response. At 0~V bias, irrespective of the drive frequency, we observe only a marginal change in the spectral character; extinction, Q-factor, and wavelength. Since the applied AC does not change the effective carrier concentration in the waveguide the spectral characteristics are unaffected with frequency. However, at a bias of +0.6~V, the AC swing ($+0.6\pm0.5$) results in increase in effective charge carriers in the waveguide. Since the voltage swing crosses the forward cut-off voltage, an exponential increase in carrier flow results in increased carriers in the waveguide. Thus decreasing the extinction ratio and Q-factor of the resonance (Fig.~\ref{fig:DiffBiasVoltage_RF1Vpp}(c)). The forward current also results in thermally-induced refractive index change hence red-shift in the resonance with frequency (Fig.~\ref{fig:DiffBiasVoltage_RF1Vpp}(d)). 

In addition to extinction and resonance wavelength shift, we also observe a change in resonance shape. The asymmetry of the cavity resonance can be evaluated using skewness factor given by,

\begin{eqnarray}
{Skewness}=\frac{E(x-\mu)^3}{\sigma^2}
\label{eq:Tn}
\end{eqnarray}

where $\mu$ is the mean of $x$, $\sigma$ is the standard deviation of $x$, and $E(t)$ represents the expected value of the quantity $t$. A negative or positive sign of the skewness indicates left skew and right skew, respectively. We investigate the effect of small-signal voltage and optical power on the skewness created by electrical and optical induced carriers. 

\begin{figure}[tb]
	\centering
	\includegraphics[width=\linewidth]{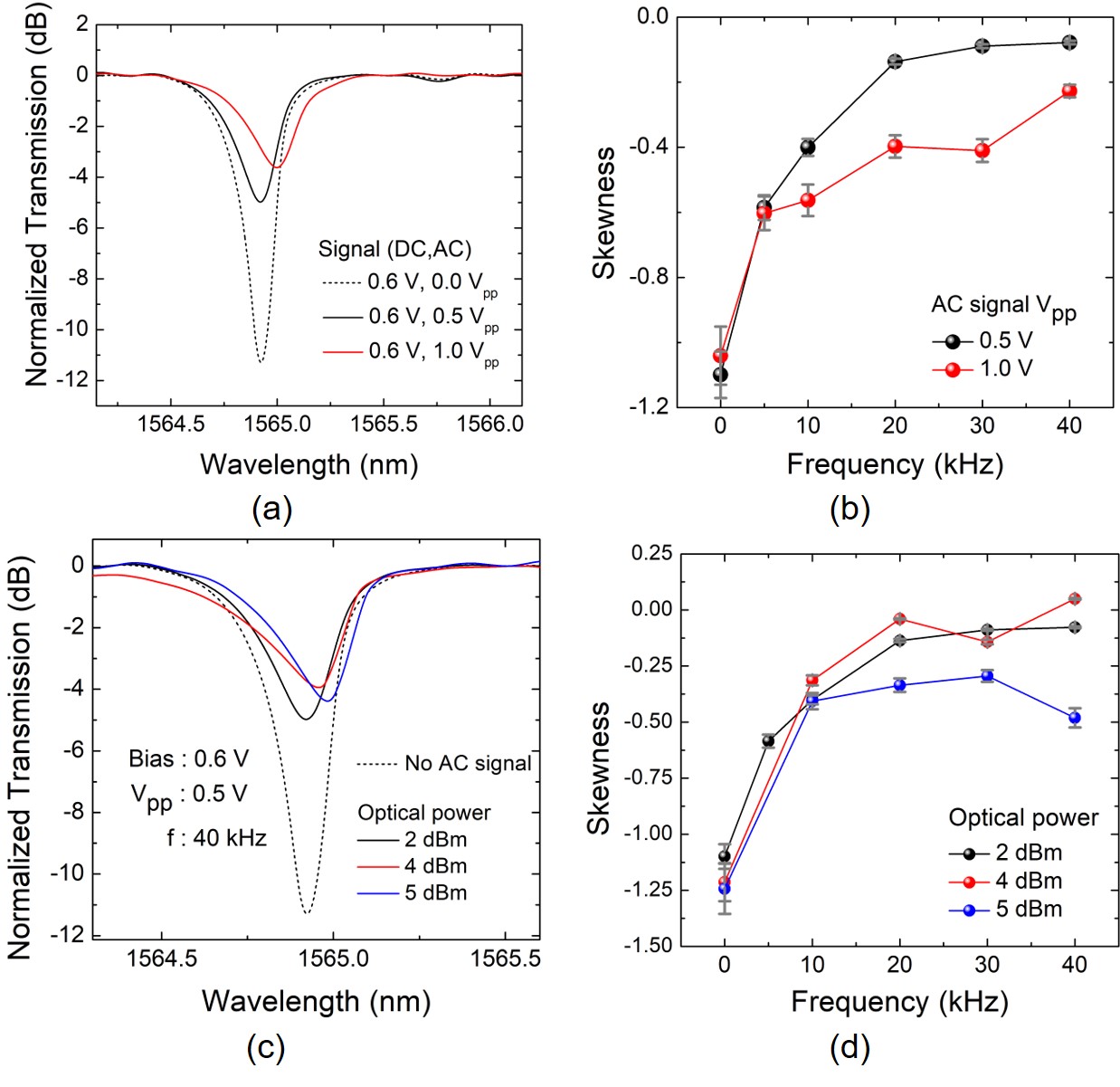}
	\caption{(a)~Transmission spectra of the ring resonator for different AC feed at bias voltage +0.6 V and optical power of +2~dBm, (b)~Skewness for increasing frequency for different AC peak-peak voltage at bias voltage +0.6 V and optical power of 2~dBm, (c)~Transmission spectra of the ring resonator for different optical power at bias voltage +0.6~V and AC swing of 0.5~V, (d)~Skewness for increasing frequency for different optical power at bias voltage +0.6~V and AC swing of 0.5~V.}
	\label{fig:RFPower_OptPower_Comp}
\end{figure}

A small signal voltage of 0.5~$V_{pp}$  and 1~$V_{pp}$ is applied with a bias of +0.6 V.  Fig.~\ref{fig:RFPower_OptPower_Comp}(a) shows the effect of small-signal voltage on the resonance. The skewness improves with increasing frequency, however, the drive voltage should be below the forward cut-off. When the voltage swing is above the forward cut-off the forward current results in excess carriers which results in suboptimal compensation of skew, which is shown in Fig.~\ref{fig:RFPower_OptPower_Comp}(b) at 1 $V_{pp}$. Furthermore, the forward current also induces thermal-optic shift. 

Finally, the effect of small-signal frequency at various optical power is investigated (Fig.~\ref{fig:RFPower_OptPower_Comp}(c)). The spectral skewness is measured at 2, 4 and 5 dBm of optical input power when the modulator is driven with a small-signal voltage of 0.5 $V_{pp}$ with a bias of +0.6 V. We observe improvement in the skewness at lower optical power $<$4 dBm, however, the skewness saturates at suboptimal skew at higher optical power~(Fig.~\ref{fig:RFPower_OptPower_Comp}(d)). As discussed in previous sections, at high optical power, the optically generated carriers saturate the waveguide beyond compensation.

\section{Conclusion}

In summary, we presented a comprehensive experimental study and analysis of the effect of electro-optic behaviour of a silicon ring modulator. The Q-factor of the ring cavity plays a significant role in the nonlinear threshold of optical power. For a high-Q cavity, the TPA and FCA induced absorption starts at low optical injection. The generated carriers can be swept by applying an electric field across the junction; however, the field is ineffective at high-optical injection. However, at moderate injection an alternating electric field could sweep the carries resulting in improved spectral response. We showed mitigation of spectral distortion of ring resonator with applied electric field across the waveguide using bias tunable AC signals. The study helps in understanding the operating point and limitation of optically generated carrier mitigation in a plasm-dispersion based silicon modulator.

\medskip
\noindent
\textbf{Funding.} We acknowledge funding support from MHRD through NIEIN project, from MeitY and DST through NNetRA project.

\medskip
\noindent
\textbf{Acknowledgement.} V.J. acknowledges Ministry of Electronics and Information Technology (MeitY), Govt. of India for the student fellowship and S.K.S acknowledges MeitY for the faculty fellowship. Authors thank Prof. Sushobhan Avasti for fruitful discussions.

\bibliographystyle{unsrt}
\bibliography{sample_bib}

\end{document}